\documentclass[%
 aip,
 amsmath,amssymb,
 reprint,%
]{revtex4-2}

\usepackage{graphicx}% Include figure files
\usepackage{dcolumn}% Align table columns on decimal point
\usepackage{bm}% bold math
\usepackage[utf8]{inputenc}
\usepackage[T1]{fontenc}
\usepackage{mathptmx}
\usepackage{etoolbox}
\usepackage{siunitx}
\usepackage{multirow}

\makeatletter
\def\@email#1#2{%
 \endgroup
 \patchcmd{\titleblock@produce}
  {\frontmatter@RRAPformat}
  {\frontmatter@RRAPformat{\produce@RRAP{*#1\href{mailto:#2}{#2}}}\frontmatter@RRAPformat}
  {}{}
}%
\makeatother
\begin{document}

\preprint{AIP/123-QED}

\newcommand{\NTTNC}{%
  NTT Nanophotonics Center, NTT, Inc., 3-1, Morinosato Wakamiya \mbox{Atsugi, 243-0198, Japan}
}
\newcommand{\NTTBRL}{%
  Basic Research Laboratories, NTT, Inc., 3-1, Morinosato Wakamiya \mbox{Atsugi, 243-0198, Japan}
}

\title{Silicon photonic optical--electrical--optical converters based on load-resistor and current-injection operation}

\author{Masaya Arahata}
\email[Corresponding author: ]{masaya.arahata@ntt.com}
\affiliation{\NTTNC}
\affiliation{\NTTBRL}

\author{Shota Kita}
\affiliation{\NTTNC}
\affiliation{\NTTBRL}

\author{Akihiko Shinya}
\affiliation{\NTTNC}
\affiliation{\NTTBRL}

\author{Hisashi Sumikura}
\affiliation{\NTTNC}
\affiliation{\NTTBRL}

\author{Masaya Notomi}
\affiliation{\NTTNC}
\affiliation{\NTTBRL}

\date{\today}% It is always \today, today,
             %  but any date may be explicitly specified

\begin{abstract}
Optical–electrical–optical (OEO) converters are key primitives for low-latency, energy-efficient photonic computing because they enable nonlinear activation and optical signal regeneration on chip. We report two monolithically integrated silicon-photonic OEO converters---load-resistor (high-speed variant) and current-injection (high-gain variant) types---fabricated at a silicon photonics foundry. Each device combines a germanium photodetector with a micro-ring modulator (MRM). The converters exhibit reconfigurable nonlinear transfer functions and measurable on-chip RF OEO gain. The RF OEO gain scales linearly with the MRM bias power, with slopes of 0.10 $\mathrm{mW}^{-1}$ (load-resistor of 10 k$\Omega$) and 1.4 $\mathrm{mW}^{-1}$ (current-injection), enabling a gain > 1 region at practical bias powers ($\sim$10 mW and $\sim$1 mW, respectively). Eye diagrams confirm clear openings up to 4 Gb/s for a high-speed load-resistor variant with a 500-$\Omega$ load. To the best of our knowledge, this is the first experimental demonstration of a monolithically integrated, foundry-fabricated silicon-photonic load-resistor type OEO converter exhibiting reconfigurable nonlinear transfer and on-chip RF OEO gain. In the carrier-injection device, the activation slope exceeds unity, yielding 3.9 dB extinction-ratio regeneration. Short-pulse measurements yield 3-dB bandwidths of 1.49 GHz, 160 MHz (load-resistor of 500 $\Omega$ and 10 k$\Omega$), and 76 MHz (current-injection), consistent with the RF data. Energy analysis shows an energy-bandwidth trade-off (RC-limited for load-resistor vs. lifetime-limited for injection) and outline routes to sub-pJ/bit operation via reduced capacitance and improved EO efficiency. These results establish silicon-photonic OEO converters as compact, foundry-compatible building blocks for scalable optoelectronic computing and optical neural networks.
\end{abstract}

\maketitle

\section{Introduction}
%電気のデメリット、光の注目、全光の問題点
Modern digital electronic systems for machine-learning and telecommunication signal processing face energy and latency bottlenecks due to charge-based switching, memory movement, and RC-limited interconnects \cite{Williams2017}, despite their facility for implementing nonlinear operations. In contrast, all-optical and photonic analog processing offers ultra-high bandwidth, low latency, and favorable energy per operation---attributes well suited to the multiply-accumulate-dominated artificial-intelligence workloads. Advances in photonic integration have therefore spurred efforts to scale optical systems across multiple degrees of freedom of light, including neuromorphic and optical neural networks (ONNs) \cite{Shen2017,Tait2017,Feldmann2021,Ohno2022,Wang2022,Pai2023,Ikeda2024}. 

%光のdisadvantage 非線形と利得
However, scaling remains challenging because (i) optical loss accumulates in interferometric meshes and feedback loops, degrading the signal-to-noise ratio as circuits grow \cite{Clements2016,Arahata2025}, and (ii) reconfigurable on-chip nonlinear elements are scarce. Without nonlinearity and on-chip regeneration (gain), multilayer optical networks collapse to a single linear transform and become intolerant to loss and drift \cite{Matuszewski2024,Fu2024}. While these functions can be achieved by converting to electronics, repeated optical-to-electrical (O/E) and analog-to-digital (A/D) conversions erode photonics' benefit in latency and energy \cite{Mustafa2024,Xia2024,Meng2020,Fu2024}.

%OEOの利点と課題
Optical--electrical--optical (OEO) converters \cite{Williamson2020}---in which a photodetector (PD) and an electro-optic modulator (EOM) are electrically coupled so that electronic nonlinearity and gain can be re-imprinted onto light---offer a promising compromise. In an OEO stage, the input optical signal is detected, processed electrically, and then used to drive an EOM on an external optical carrier. No A/D conversion is required; however, the O/E step typically dominates energy and latency, and conventional designs add a transimpedance amplifier (TIA) after the PD \cite{Ashtiani2022}, which further increases the power consumption. Many prior methods also interconnect discrete, fiber-coupled components \cite{Krehlik2021}. Nanophotonic integration mitigates these penalties: device miniaturization reduces EO switching energy, and pairing a nanoscale, ultralow-capacitance PD with a resistive load (load-resistor type) improves O/E efficiency without requiring a TIA. By avoiding optical amplification---and thus amplified spontaneous emission noise---the approach is naturally compatible with complementary metal-oxide-semiconductor technology \cite{nozaki2018IEEE}. We previously demonstrated an ultra-low-power InP photonic-crystal OEO converter operating at 10 Gb/s with 1-fJ optical energy, enabled by nanoscale PD/EOM capacitive scaling \cite{nozaki2019}. Yet, heterogeneous co-integration with silicon photonics remains complex, motivating monolithic silicon-photonic OEO converters \cite{Hector2022} that deliver on-chip gain and reconfigurable nonlinear activation within foundry processes. Moreover, because OEO converters comprising only a PD and an EOM can exhibit far smaller capacitance than photonics-electronics co-integration based on co-packaged optics, continued nanophotonics scaling on silicon is expected to further reduce drive energy and footprint, thereby strengthening OEO's role as a building block for large-scale optical computing.

%これまでのSiPhOEOの状況、負荷抵抗型と電流注入型の展望
Recent work has demonstrated monolithic silicon-photonic ONN circuits (feedforward and recurrent) using current-injection OEO stages in which the photocurrent directly drives an EOM \cite{Bandyopadhyay2024,Wu2025,Wu2025-2,Wu2025-3}. Load-resistor and current-injection OEOs present complementary trade-offs: the latter's bandwidth is fundamentally limited by carrier-recombination lifetime (typically $\sim100$ MHz), whereas the former is RC-limited and accelerates with reduced device capacitance. Current-injection designs generally require lower drive voltage and energy. In both schemes, capacitance minimization is central to lowering the energy, and in the load-resistor type, it also directly increases speed. 

%これまでのSiPhOEOの状況
Despite this progress, a clear research gap remains. Reports of integrable OEO converters for optical computing systems are limited: for the load-resistor type, only an InP photonic-crystal OEO converter has been reported \cite{nozaki2019}, and to our knowledge, no silicon-photonic load-resistor OEO converters integrated with optical circuits have been demonstrated. For the current-injection type, several silicon-photonic OEO converters integrated with optical meshes have been reported \cite{Bandyopadhyay2024,Wu2025,Wu2025-2,Wu2025-3}, but the OEO stages mainly served as nonlinear activations and their on-chip RF gain has not been quantified.

%本研究の目的
In this work, we report monolithically integrated silicon-photonic OEO converters in two architectures: a load-resistor design and a current-injection design. Both converters exhibit reconfigurable nonlinear transfer functions and measurable on-chip RF OEO gain. The load-resistor device achieves RF gain at an operating speed of 200 Mb/s with the load resistance $R_\mathrm{load}$ of 10 k$\Omega$, and a higher-speed variant with $R_\mathrm{load}=500$ $\Omega$ shows clear eye openings up to 4 Gb/s, illustrating the expected RC trade-off between gain efficiency and bandwidth. To the best of our knowledge, this is the first experimental demonstration of a monolithically integrated silicon-photonic load-resistor OEO converter without a TIA showing reconfigurable nonlinear transfer and measurable on-chip RF OEO gain. The current-injection device provides a larger RF OEO gain and strong extinction-ratio (ER) regeneration due to high EO efficiency under forward bias while its RF speed is limited to about 100 Mb/s, primarily by the carrier recombination lifetime. Short-pulse measurements independently give 3-dB bandwidths of 1.49 GHz, 160 MHz (load-resistor, $R_\mathrm{load}=500$ $\Omega$ and 10 k$\Omega$), and 76 MHz (current-injection), consistent with the RF data. In addition to RF metrics, we present a unified energy per bit model and measurements for both OEO architectures. The energy per bit analysis quantitatively maps the speed–energy trade-offs and provides clear designs for future scaling. These results give two complementary options: use the depletion-mode design for high-throughput (RC-limited) stages, and use the injection-mode design when the higher gain or ER regeneration is needed---both of which can be implemented monolithically on silicon photonics.

In Section II of this paper, we present the device architectures and experimental characterization of both OEO converters. Section III analyzes the mechanisms governing OEO gain and bandwidth (including the trade-off between load resistance and speed), compares our results with prior work, and evaluates the energy per bit of the load-resistor and current-injection OEO converters. Section IV summarizes the main findings and concludes the paper.  

\section{Results}

\subsection{Device details and experimental setup}

\begin{figure*}
  \begin{center}
  \includegraphics[width=7.1in]{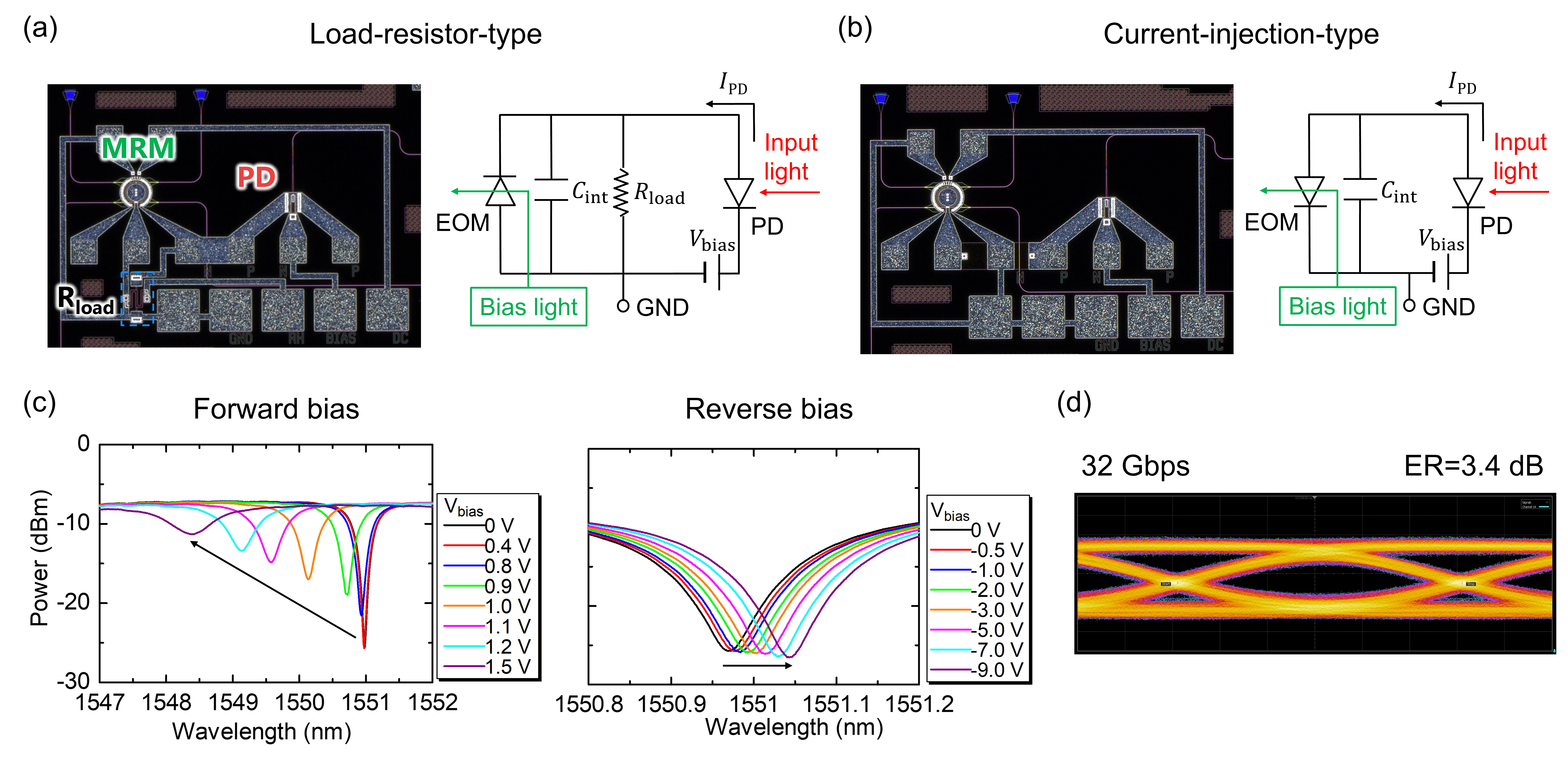}\\
  \caption{Micrographs and equivalent circuit diagrams of (a) load-resistor-type and (b) current-injection-type OEO converters. In the load-resistor-type converter, the p-doped region of the PD is connected to the n-doped region of the MRM's p-n-junction phase shifter, whereas in the current-injection-type converter, it is connected to the p-doped region. (c) Optical spectra of a single MRM under forward and reverse bias, corresponding to carrier-injection and carrier-depletion modes, respectively. (d) Eye diagram of the RF operation of the MRM when a 32 Gb/s NRZ voltage signal ($V_\mathrm{pp}=1.5$ V and $V_\mathrm{off}=-2.7$ V) is applied to its phase shifter, showing a clear opening with an ER of 3.4 dB. $C_\mathrm{tot}$: total capacitance contributed by the PD and the EOM. $R_\mathrm{load}$: load resistance. $I_\mathrm{PD}$: photocurrent. $V_\mathrm{bias}$: bias voltage. GND: electrical ground.}\label{fig_spectra}
  \end{center}
\end{figure*}

First, we describe the device details of the load-resistor-type and current-injection-type OEO converters. Figure \ref{fig_spectra}(a) and (b) show micrographs and equivalent circuit diagrams of load-resistor-type and current-injection-type devices, respectively. The chips were monolithically fabricated at Advanced Micro Foundry (AMF) on a silicon photonics platform. The optical loss at an edge coupler is approximately 3 dB. In both designs, the EOM is a micro-ring modulator (MRM) with a typical cavity quality factor of $Q\sim$4000. In the load-resistor-type converter, the p-doped region of the germanium PD is connected to the n-doped region of the MRM's p-n-junction phase shifter, whereas in the current-injection-type converter, it is connected to the p-doped region. The load resistor is implemented through p-doping of the silicon waveguide. Under reverse bias $V_{\rm{bias}}$, an input light on the PD generates a photocurrent flowing through $R_{\rm{load}}$, generating a voltage that drives the EOM in carrier-depletion mode for high-speed operation. The operation speed is primarily limited by the device RC time constant, $R_{\rm{load}}C_{\rm{tot}}$, where $C_{\rm{tot}}$ is the total capacitance contributed by the PD and the EOM, including parasitic capacitances from electrical pads. In the current-injection-type converter, the photocurrent flows directly into the EOM, which is therefore operated in carrier-injection mode for high gain. Unless otherwise noted, the reported ``PD input power'' and ``EOM output power'' represent the on-chip optical power, taking into account edge-coupler losses and assuming 3.2 dB/facet at a wavelength of 1520 nm and 3.0 dB/facet near 1550 nm, as well as a waveguide propagation loss of 0.1 dB/mm, according to the AMF process design kit.

Figure \ref{fig_spectra}(c) shows the spectra of a stand-alone MRM under forward and reverse bias. The resonance shift is much larger under forward bias than under reverse bias, corresponding to carrier-injection and carrier-depletion operation, respectively, indicating that the injection-type modulator requires a substantially lower operating voltage than its depletion-type counterpart. From Fig. \ref{fig_spectra}(c), the resonance-shift efficiencies are $d\lambda/dV$ of $\simeq1.7$ nm/V (injection) and $\simeq8.9$ pm/V (depletion). With $Q\simeq4000$ at 1550 nm, the voltage to shift by half the linewidth is roughly $\{\lambda/(2Q)\}/(d\lambda/dV)$, i.e., $\sim0.1$ V (injection) and $\sim22$ V (depletion). Figure \ref{fig_spectra}(d) shows the eye diagram of the RF operation of the MRM when a 32 Gb/s non-return-to-zero (NRZ) voltage signal with peak-to-peak voltage $V_\mathrm{pp}=1.5$ V and DC offset $V_\mathrm{off}=-2.7$ V is applied to its phase shifter. The eye pattern is clearly open at this data rate with an ER of 3.4 dB.

\begin{figure*}
  \begin{center}
  \includegraphics[width=6.5in]{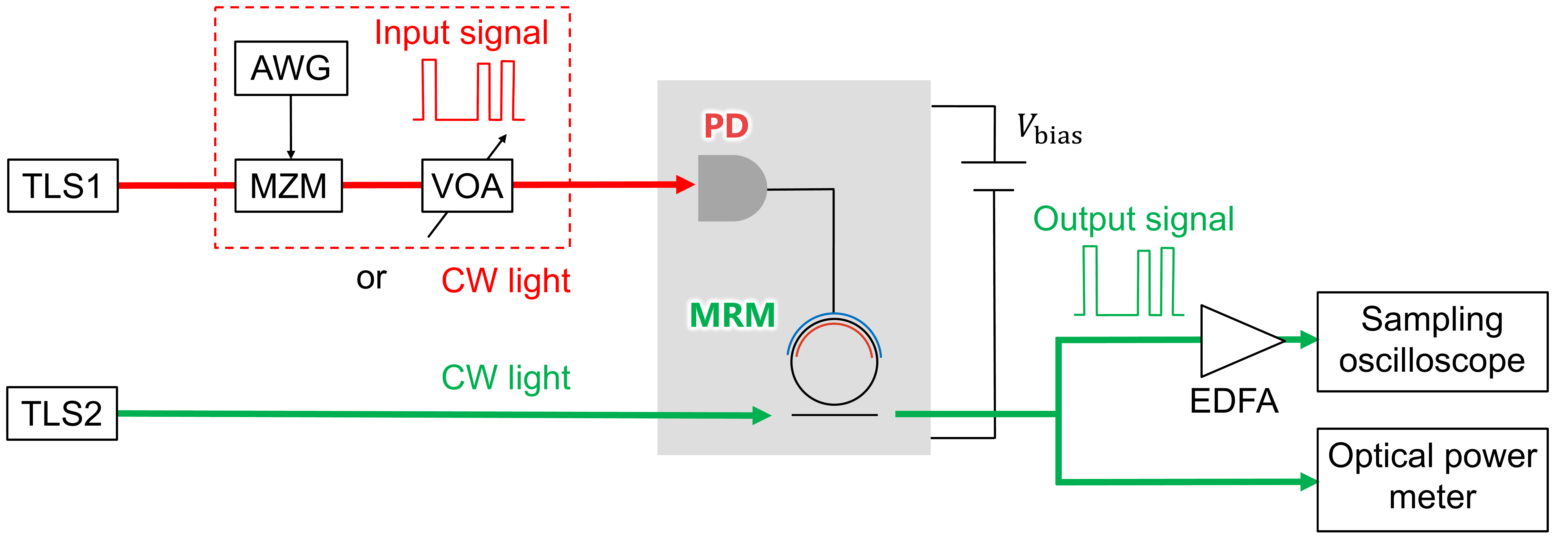}\\
  \caption{Experimental setup for evaluating the OEO converters. The PD is driven either by a CW optical input for nonlinear activation measurements or by an RF-modulated optical input with an external modulator; a CW bias light is injected into the MRM. The output is amplified by an EDFA and recorded with a sampling oscilloscope and an optical power meter. TLS: tunable laser source. AWG: arbitrary waveform generator. MZM: Mach-Zehnder modulator. VOA: variable optical attenuator. EDFA: erbium-doped fiber amplifier.}\label{fig_ES}
  \end{center}
\end{figure*}

Figure \ref{fig_ES} illustrates the experimental setup for evaluating the performance of the OEO converters. For the evaluation of nonlinear activation functions and RF OEO gain, a continuous-wave (CW) input signal or an RF input signal is incident on the PD, respectively. The RF input signal is generated by an arbitrary waveform generator (AWG) and an EOM. The CW bias light is launched into the MRM, and then the output signal is subsequently amplified by an erbium-doped fiber amplifier (EDFA) and measured with a sampling oscilloscope and an optical power meter. The EDFA is not used during the RF OEO-gain measurements. 

\subsection{Load-resistor-type OEO converter}

\begin{figure*}
  \begin{center}
  \includegraphics[width=6.5in]{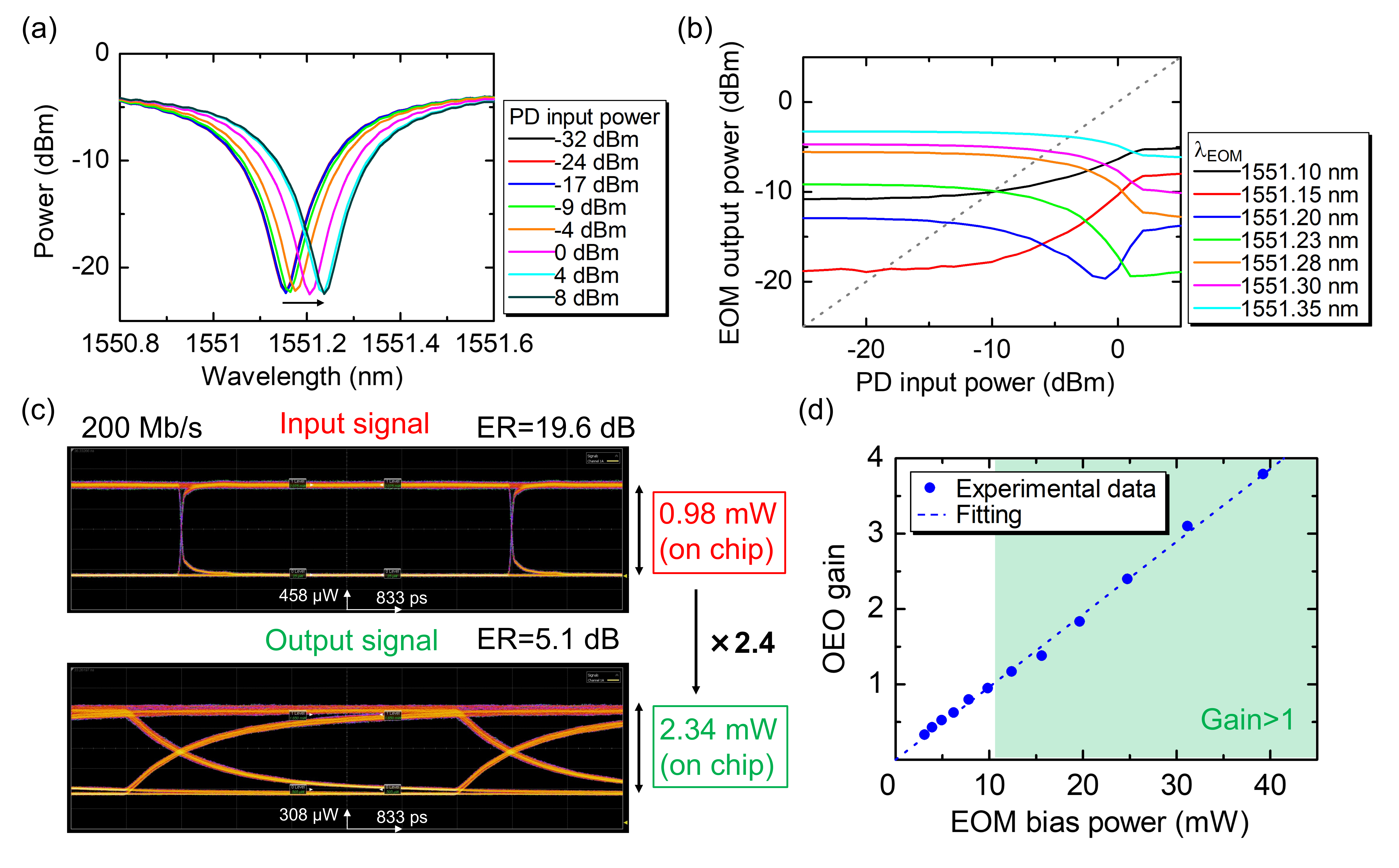}\\
  \caption{Evaluation results of the load-resistor-type OEO converter with $R_{\rm{load}}=10$ k$\Omega$ under $V_\mathrm{bias}=-8$ V. (a) Spectra of the load-resistor-type OEO converter under varying CW input powers incident on the PD, including the loss at the edge coupler. (b) Nonlinear activation functions at different operating wavelengths with $P_\mathrm{bias}=0$ dBm; $\lambda_{\mathrm{EOM}}$ denotes the wavelength of the EOM bias light. (c) Eye diagrams for RF PD input NRZ signal (top panel) at 200 Mb/s with an ER of 19.6 dB and the corresponding EOM output signal (bottom panel) obtained with $P_\mathrm{bias}=14$ dBm (25 mW) at an operating wavelength of 1551.48 nm. The ER of the EOM output signal is 5.1 dB. (d) Dependence of the RF OEO gain on the EOM bias light power at 200 Mb/s, when the ER of the EOM output signal is fixed to 5 dB, showing a linear increase with increasing bias light power.}\label{fig_NTT10kohm}
  \end{center}
\end{figure*}

\begin{figure*}
  \begin{center}
  \includegraphics[width=6.0in]{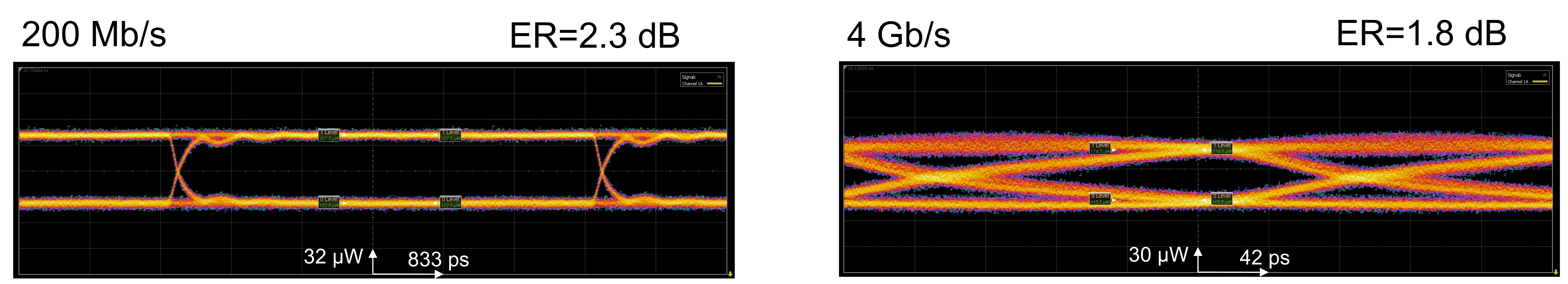}\\
  \caption{Eye diagrams of the load-resistor-type OEO converter with $R_{\rm{load}}=500$ $\Omega$ for the RF-modulated optical signals at 200 Mb/s (left) and 4 Gb/s (right). $P_{\rm{bias}}=3$ dBm and $V_{\rm{bias}}=-5$ V. The output optical signals are amplified by an EDFA. The ERs are 2.3 and 1.8 dB, respectively.}\label{fig_NTT500ohm}
  \end{center}
\end{figure*}

We first study the $R_\mathrm{load}=10$ k$\Omega$ device (gain-focused), then a speed-optimized variant with $R_\mathrm{load}=500$ $\Omega$ (Figs. \ref{fig_NTT10kohm} and \ref{fig_NTT500ohm}). Figure \ref{fig_NTT10kohm}(a) shows the spectra of the output from the converter with $R_{\rm{load}}=10$ k$\Omega$ under varying CW input powers incident on the PD, including the loss at the edge coupler (assumed 3.2 dB/facet at a wavelength of 1520 nm according to the AMF process design kit) when $V_\mathrm{bias}=-8$ V. As the input power increases, the spectra shift toward longer wavelengths in carrier-depletion mode via the load voltage. The maximum wavelength shift is approximately 0.05 nm when the PD input power reaches 8 dBm. Figure \ref{fig_NTT10kohm}(b) shows the nonlinear activation functions at different operating wavelengths with $P_\mathrm{bias}=0$ dBm, corresponding to the relations between the PD input power and the EOM output power on the chip including the loss at the edge coupler (assumed 3.0 dB/facet around wavelengths of 1550 nm). Distinct nonlinear functions are observed depending on the operating wavelengths. The maximum ER of the EOM output power is about 10 dB at an operating wavelength of 1551.15 nm. It should be noted that the plateau observed for PD input powers above 0 dBm is attributed to PD saturation. These results clearly demonstrate that the OEO converter is capable of representing complex nonlinear activation functions for ONNs, where reconfigurability of the functions is essential. 

To demonstrate the amplification of the RF input signal by OEO conversion, we evaluated the RF OEO gain $G$ including the coupling loss at the edge coupler. Figure \ref{fig_NTT10kohm}(c) shows the eye diagrams for the RF PD input NRZ signal at 200 Mb/s with an ER of 19.6 dB and the corresponding EOM output signal obtained with $P_\mathrm{bias}=14$ dBm (25 mW) at an operating wavelength of 1551.48 nm under $V_\mathrm{bias}=-8$ V. The ER of the EOM output signal is 5.1 dB. From the RF signal amplitudes, the RF OEO gain $G$ was estimated to be 2.4. Figure \ref{fig_NTT10kohm}(d) presents the dependence of $G$ on $P_\mathrm{bias}$ when the ER of the EOM output signal is fixed to 5 dB, showing a linear increase with increasing bias light power. Linear fitting yields a slope of 0.10 $\mathrm{mW}^{-1}$. These results indicate that the gain region ($G>1$) is achieved when $P_\mathrm{bias}$ exceeds 10 mW. The maximum $G$ of 3.8 obtained with our OEO converter surpasses the reported value of 2.3$\pm$0.3 \cite{nozaki2019} at 200 Mb/s  for a load-resistor-type InP photonic crystal cavity OEO converter, in which the gain is limited by the absorption of the EOM bias light within the modulator. Therefore, we demonstrate that the OEO converter can operate in the gain regime when sufficient EOM bias light power is applied on a silicon photonics chip. 

To verify the high-speed operation of the load-resistor-type OEO converter, we characterized the device with $R_{\rm{load}}=500$ $\Omega$. Figure \ref{fig_NTT500ohm} shows eye diagrams for RF-modulated optical inputs at 200 Mb/s (left) and 4 Gb/s (right) where $P_{\rm{bias}}=3$ dBm and $V_{\rm{bias}}=-5$ V. The output optical signals are amplified by an EDFA. The ERs are 2.3 and 1.8 dB, respectively. These results demonstrate broadband operation and indicate that the achievable data rate can be widely tuned by adjusting the load resistance.

\subsection{Current-injection-type OEO converter}

\begin{figure*}
  \begin{center}
  \includegraphics[width=6.5in]{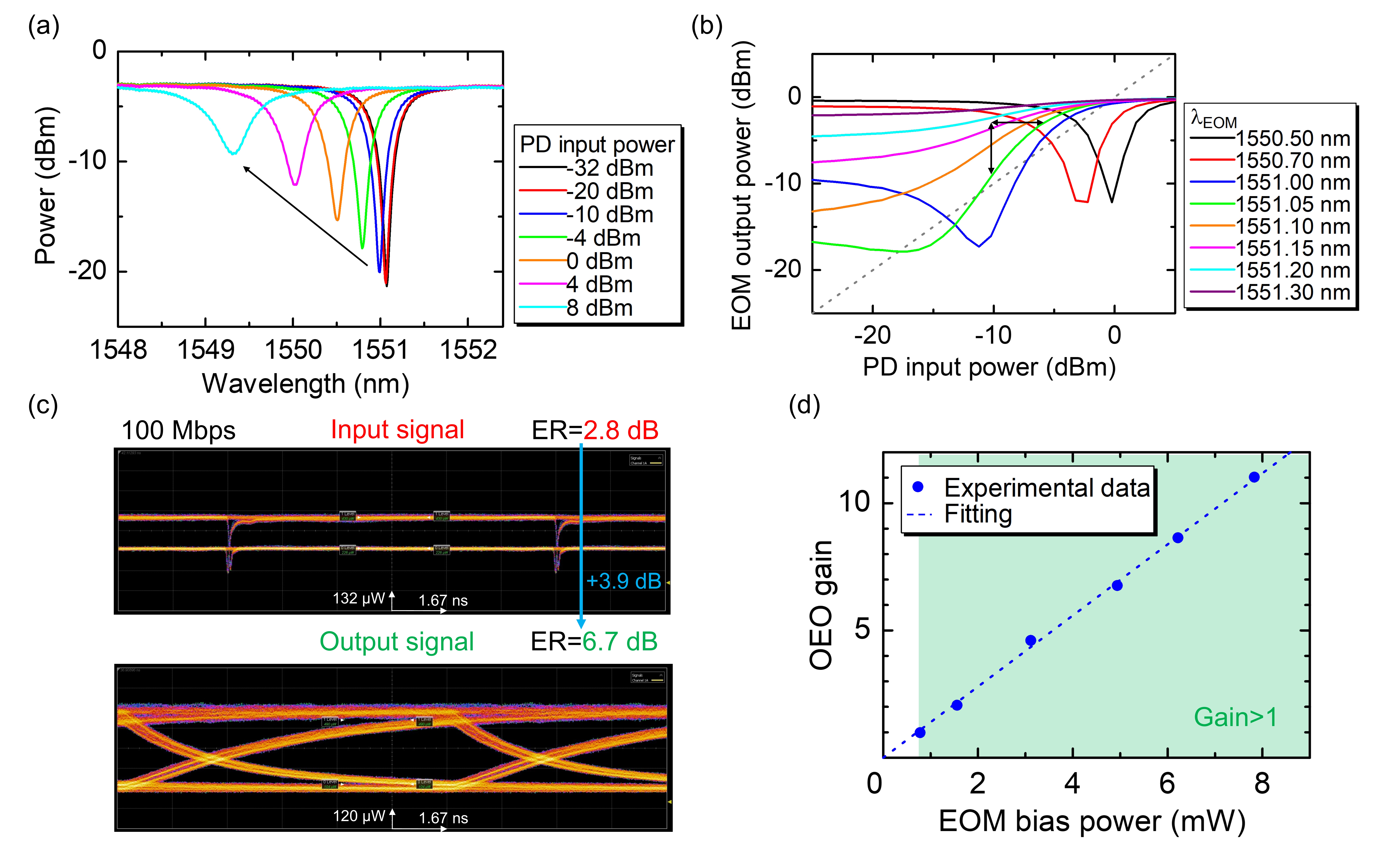}\\
  \caption{Evaluation results of the current-injection-type OEO converter under $V_{\rm{bias}}=2$ V. (a) Transmission spectra under different CW optical powers incident on the PD (power values include edge-coupler loss). (b) Nonlinear activation functions at several operating wavelengths with $P_\mathrm{bias}=0$ dBm, revealing distinct response curves. The maximum ER of the EOM output is about 18 dB at the operating wavelength of 1551.05 nm. (c) Eye diagrams for the PD input NRZ signal at 100 Mb/s with ER $=$ 2.8 dB and the corresponding EOM output obtained with $P_\mathrm{bias}=0$ dBm (1 mW) at the operating wavelength of 1551.05 nm. The output optical signal is amplified by an EDFA. The output ER is 6.7 dB, indicating an ER improvement (regeneration) of 3.9 dB, which is in good agreement with the relation between the estimated PD input power and the EOM output power obtained from the DC characteristics in (b) (arrows). (d) Dependence of the RF OEO gain on the EOM bias light power at 200 Mb/s, showing a linear increase; a linear fit gives a slope of 1.4 $\mathrm{mW}^{-1}$.}\label{fig_MIT}
  \end{center}
\end{figure*}

To compare the performance of the OEO converters, we evaluated the current-injection-type OEO converter. The measurement procedure was the same as that used for the load-resistor-type OEO converter in the previous section. Figure \ref{fig_MIT} summarizes the measurement results. Figure \ref{fig_MIT}(a) shows transmission spectra under different CW optical powers incident on the PD (power values include edge-coupler loss) under $V_{\rm{bias}}=2$ V. As the input power increases, the resonance exhibits a blue shift due to a photocurrent-driven carrier injection. The maximum shift is about 1.7 nm at a PD input power of 8 dBm. Figure \ref{fig_MIT}(b) shows the nonlinear activation functions at several operating wavelengths with $P_\mathrm{bias}=0$ dBm, revealing distinct response curves. The maximum ER of the EOM output is about 18 dB at the operating wavelength of 1551.05 nm. At this wavelength, the slope of the nonlinear function exceeds 1, leading to ER regeneration (recovery of a low-ER input). These results clearly demonstrate that the OEO converter is capable of representing complex nonlinear activation functions for ONNs. Moreover, its high-nonlinear transfer makes it useful as an optical repeater with ER regeneration \cite{Matthew2006} and an optical thresholder \cite{Huang2019} without a semiconductor optical amplifier (SOA) or TIA for optical computing and telecommunication systems.

To confirm that the RF OEO gain is higher in carrier-injection mode than in carrier-depletion mode, we characterized $G$ and confirmed clear eye diagrams up to 100 Mb/s under $V_{\rm{bias}}=2$ V. Figure \ref{fig_MIT}(c) shows eye diagrams for the PD input NRZ signal at 100 Mb/s with ER $=$ 2.8 dB and the corresponding EOM output obtained with $P_\mathrm{bias}=0$ dBm (1 mW) at the operating wavelength of 1551.05 nm. The output optical signal is amplified by an EDFA operated in the linear region. The output ER is 6.7 dB, indicating an ER improvement (regeneration) of 3.9 dB. This behavior agrees with the relation between the estimated PD input power and the EOM output power obtained from the DC characteristics in Fig. \ref{fig_MIT}(b) (arrows). Figure \ref{fig_MIT}(d) plots the dependence of $G$ on $P_\mathrm{bias}$ at 200 Mb/s, showing a linear increase; a linear fit gives a slope of 1.4 $\mathrm{mW}^{-1}$, which is more than 10 times larger than that of the load-resistor-type OEO converter. The gain region ($G>1$) is reached for $P_{\mathrm{bias}}\geq$ 1 mW. These results demonstrate high-gain operation and ER regeneration, both of which are advantageous for processing low-power and low-ER optical signals in optical applications.

\subsection{Short-pulse response}

\begin{figure}
  \begin{center}
  \includegraphics[width=2.5in]{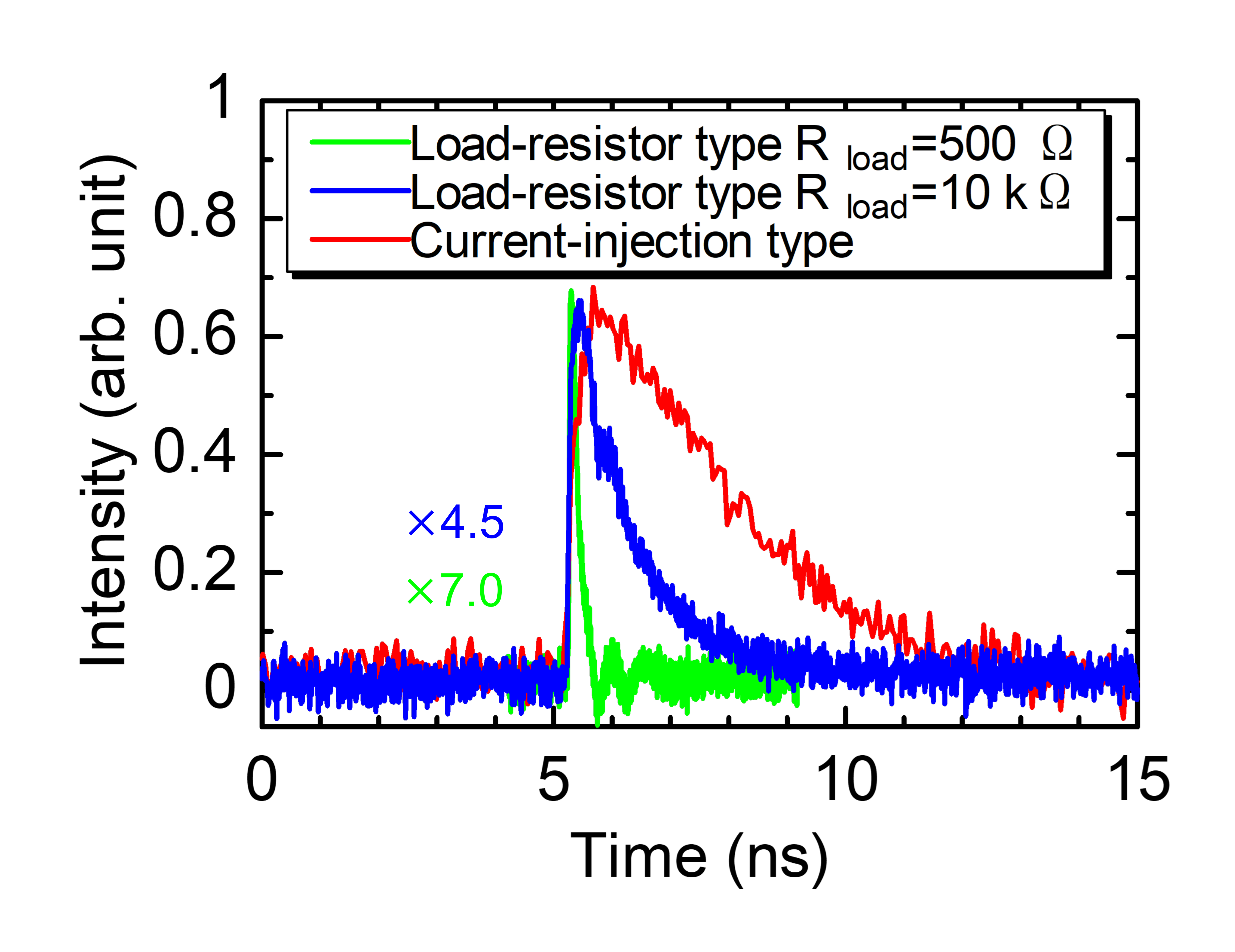}\\
  \caption{Short-pulse (impulse) responses of the EOM output for load-resistor-type OEO converters with $R_{\mathrm{load}}=500$ $\Omega$ and 10 k$\Omega$ , and for a current-injection-type OEO converter. The background level of the output power was subtracted for all traces. For visual clarity, the responses of the load-resistor-type devices with $R_{\mathrm{load}}=500$ $\Omega$ and 10 k$\Omega$ were scaled by factors of 7.0 and 4.5, respectively.}\label{fig_pulse}
  \end{center}
\end{figure}

To characterize the operation speed of the OEO converters, we measured the short-pulse (impulse) response of the load-resistor-type with $R_{\mathrm{load}}=500$ $\Omega$ and 10 k$\Omega$ and the current-injection-type device. A 22-ps-wide optical pulse was injected into the PD, and the resulting EOM output was recorded. Figure \ref{fig_pulse} shows the transient responses of the EOM output power. Note that the background level of the output power was subtracted for all traces. For visual clarity, the responses of the load-resistor-type devices with $R_{\mathrm{load}}=500$ $\Omega$ and 10 k$\Omega$ were scaled by factors of 7.0 and 4.5, respectively. The rise times are very short for both devices, whereas the fall time of the load-resistor-type converter is shorter than that of the current-injection-type converter. The time constants were extracted by fitting the traces with $y=y_{0}+A\{\mathrm{exp}{(-(t-t_{\mathrm{th}})/\tau_{f})}-\mathrm{exp}{((t-t_{\mathrm{th}})/\tau_{r})\}H(t-t_{\mathrm{th}})}$, where $y_{0}$ is a baseline, $A$ is an amplitude, $t_{\mathrm{th}}$ is the threshold (excitation onset) time, $\tau_{r, f}$ are the rise and fall time constants, respectively, and $H(t)$ is the Heaviside step function. From the fits, we obtain $\tau_{f}=107$ ps, 1.00 ns, and 2.1 ns for the load-resistor-type with $R_{\mathrm{load}}=500$ $\Omega$ and 10 k$\Omega$ and the current-injection-type OEO converters, respectively, corresponding to single-pole 3-dB bandwidths $f_{3\,\mathrm{dB}}\simeq 1/(2\pi\tau_{f})$ of 1.49 GHz, 160 MHz, and 76 MHz. These values are consistent with the independently measured RF operation speeds in the previous sections.

\section{Discussion}

\subsection{Bandwidth and gain improvement of load-resistor OEO converter}

\begin{table*}
\caption{\label{OEOtable} Measured performance of the OEO converters. $G$ is the RF OEO gain and $P_{\rm{bias}}$ is the optical bias power applied to the EOM; the ratio $G/P_{\mathrm{bias}}$ is reported in $\mathrm{mW}^{-1}$. $\tau_{RC}$ is the RC time constant. $f_\mathrm{mod}$ is the NRZ symbol rate, $f_{3\,\mathrm{dB}}$ is the 3-dB bandwidth. ER denotes the extinction ratio. $E_{\mathrm{opt\_PD}}$ is the input optical energy per bit for the PD. $E_{\mathrm{elect}}$ is the electrical energy associated with the photocurrent dissipation, i.e., the work done by the PD bias source. $E_{\mathrm{opt\_EOM}}$ is the energy per bit of optical bias to the EOM. $E_{\mathrm{ch}}$ is the dynamic charge/discharge energy of the EOM. Values in parentheses are calculated using the intrinsic capacitances of the PD and EOM, excluding pad capacitance.}
\centering
{\renewcommand\arraystretch{1}
\scalebox{1.03}{
\begin{tabular}{c|c|c|c|c|c|c|c|c|c|c}
\hline
 OEO type & $R_{\rm{load}}$ & $f_\mathrm{mod}$ & ER (dB) & $G$/$P_{\rm{bias}}$ ($\mathrm{mW}^{-1}$) & $\tau_{RC}$ & $f_{3\,\mathrm{dB}}$ & $E_{\mathrm{opt\_PD}}$ & $E_{\mathrm{elect}}$ & $E_{\mathrm{opt\_EOM}}$ & $E_{\mathrm{ch}}$ \\
\hline

\hline
\multirow{2}{*}{Load-resistor} & 500 $\Omega$ & 4 Gb/s & 1.8 & -- & 107 ps & 1.49 GHz & 62.5 fJ/bit & 112.5 fJ/bit & 500 fJ/bit & 6.25 (1.13) fJ/bit \\
& 10 k$\Omega$ & 200 Mb/s & 5.1 & 0.10 & 1.00 ns & 160 MHz & 1.25 pJ/bit & 2.25 pJ/bit & 50 pJ/bit & 625 (113) fJ/bit \\

\hline
Current-injection & -- & 100 Mb/s & 2.8$\rightarrow$6.7 & 1.4 & 2.1 ns & 76 MHz & 2.5 pJ/bit & 4.5 pJ/bit & 10 pJ/bit & 50 fJ/bit \\

\hline

\end{tabular}
}}
\end{table*}

In this section, we compare the evaluation results of the OEO converters and discuss the performance improvements. First, we summarize the key metrics in Table \ref{OEOtable}, which highlights a clear gain-bandwidth trade-off between the two OEO architectures. The RF OEO gain $G$ is defined as the on-chip RF amplitude ratio between EOM output and photodetector (PD) input. The gain efficiency is $G/P_\mathrm{bias}$, where $P_\mathrm{bias}$ is the optical bias power applied to the EOM. The load-resistor (depletion-mode) device supports higher-speed operation, reaching 4 Gb/s with $R_\mathrm{load}=500$ $\Omega$. Increasing $R_\mathrm{load}$ to 10 k$\Omega$ boosts the measured gain efficiency of 0.10 $\mathrm{mW^{-1}}$ at the cost of the speed of 200 Mb/s. In contrast, the current-injection-type device provides a much larger gain efficiency of 1.4 $\mathrm{mW^{-1}}$ and strong ER regeneration, while its RF speed is limited to $\sim100$ Mb/s by the carrier-lifetime constraints.

To further improve the bandwidth of the load-resistor-type OEO converter, it is essential to reduce the total capacitance $C_\mathrm{tot}$ of the PD and EOM. From the short-pulse response measurement, we obtained $\tau_{f}\sim$1.0 ns at $R_\mathrm{load}=10$ k$\Omega$, implying $C_\mathrm{tot}=\tau_{f}/R_\mathrm{load}\sim$100 fF (including pad capacitance). To push $f_\mathrm{3\,dB}$ beyond 1 GHz with the same $R_\mathrm{load}$, one would need $\tau_{f}<$160 ps, i.e., $C_\mathrm{tot}<$16 fF. It requires (i) shrinking the pad/interconnect parasitic capacitance, (ii) lowering the junction capacitance via p-n geometry and doping optimization, and (iii) shrinking the active device area (smaller-radius rings). In parallel, lowering $R_\mathrm{load}$ (e.g., to 1--2 k$\Omega$) increases the RC bandwidth at the cost of OEO gain efficiency, which is consistent with our observed gain-speed trade-off. 

By utilizing photonic-crystal cavity PDs and modulators, we can decrease $C_\mathrm{tot}$ down to the few-fF range and reduce a $\pi$ shift voltage, enabling multi-Gb/s operation; our earlier InP device reaches 10 Gb/s \cite{nozaki2019}. With careful parasitic control, a silicon-photonics implementation could approach similar speeds while keeping the additional-amplifier-free OEO architecture. 

Raising the MRM's modulation efficiency increases the OEO gain and the steepness and shape of the nonlinear activation. A higher loaded $Q$ increases the voltage-to-intensity conversion efficiency but may narrow the optical bandwidth. For our current $Q\sim4000$, the photon lifetime is only a few picoseconds at 1550 nm, which is well below the electrical RC limit, leaving room to increase $Q$ without constraining GHz-class operation. Adjusting the coupling toward near-critical coupling \cite{Christopher2004} and using coupled-resonator \cite{Li2008,Hu2012}, MZI-assisted rings \cite{Gutierrez2012}, and microdisk \cite{Soref2020} can further steepen and deepen the transfer function while preserving ER. In short, co-optimizing $C_\mathrm{tot}$, $R_\mathrm{load}$, and loaded $Q$ offers a clear path to >1 Gb/s depletion-mode bandwidth with a stronger RF OEO gain and ER regeneration, all without optical amplifiers.

\subsection{Energy consumption of OEO converters}
To assess suitability for optical computing systems, we compare the energy per bit of load-resistor and current-injection OEO converters (Table \ref{OEOtable}). We define the input optical energy per bit for the PD as $E_{\mathrm{opt\_PD}}=P_{\mathrm{PD}}/4f_{\mathrm{mod}}$ \cite{nozaki2019}, where $P_{\mathrm{PD}}$ is the optical power corresponding to one level of the NRZ signal and $f_{\mathrm{mod}}$ is the NRZ symbol rate. The electrical energy associated with the photocurrent dissipation---i.e., the work done by the PD bias source---is $E_{\mathrm{elect}}$ and is given by $E_{\mathrm{opt\_PD}}\eta_{\mathrm{PD}}|V_{\mathrm{PD}}|$, where $\eta_{\mathrm{PD}}$ is the PD responsivity and $V_{\mathrm{PD}}$ is the reverse bias voltage. The total energy per bit is 
\begin{equation}\label{OEO_energy}
E_{\mathrm{OEO,bit}}=E_{\mathrm{opt\_PD}}+E_{\mathrm{elect}}+E_{\mathrm{opt\_EOM}}+E_{\mathrm{ch}},
\end{equation}
where $E_{\mathrm{opt\_EOM}}=P_{\mathrm{bias}}/f_{\mathrm{mod}}$ is set by the EOM optical bias power $P_{\mathrm{bias}}$, and $E_{\mathrm{ch}}$ is the dynamic charge/discharge energy of the EOM.

\paragraph{Load-resistor type}
In this case, the EOM is voltage-driven in carrier-depletion mode, giving
\begin{equation}\label{load-resistor_charge}
E_{\mathrm{ch}}^{\mathrm{load}}=C_{\mathrm{tot}} (V_{\mathrm{EOM}}^{\mathrm{load}})^{2}/4.
\end{equation}
With $R_{\mathrm{load}}=10$ k$\Omega$ at $f_{\mathrm{mod}}=200$ Mb/s, we obtain $E_{\mathrm{opt\_PD}}=1.25$ pJ/bit and $E_{\mathrm{elect}}=2.25$ pJ/bit for $P_{\mathrm{PD}}=1.0$ mW, $\eta_{\mathrm{PD}}=0.9$ A/W, and $V_{\mathrm{PD}}=-2$ V. With $P_{\mathrm{bias}}\sim10$ mW (corresponding to RF OEO gain $\sim1$) and $f_{\mathrm{mod}}=200$ Mb/s, $E_{\mathrm{opt\_EOM}}=50$ pJ/bit. Using $C_{\mathrm{tot}}\sim100$ fF (this work) and $V_{\mathrm{EOM}}^{\mathrm{load}}\sim5$ V, we estimate $E_{\mathrm{ch}}^{\mathrm{load}}=625$ fJ/bit. When the pad area can be shrunk and ultimately eliminated in optical computing applications such as ONN circuits, $C_{\mathrm{tot}}$ is reduced to $\sim$5 fF (PD) + 13 fF (EOM) \cite{Yuan2022}, i.e., $\sim$18 fF, resulting in $E_{\mathrm{ch}}^{\mathrm{load}}\sim113$ fJ/bit. With further EO-efficiency improvements and reduced device capacitance, $P_{\mathrm{bias}}$ can be lowered even more, enabling $E_{\mathrm{opt\_EOM}}<1$ pJ/bit and $E_{\mathrm{ch}}^{\mathrm{load}}<100$ fJ/bit in principle. If heterogeneous co-integration with silicon photonics advances, operation at 10 Gb/s with a device capacitance of 2 fF using an InP photonic-crystal nanocavity \cite{nozaki2019} would yield $E_{\mathrm{opt\_EOM}}\sim10$ fJ/bit and $E_{\mathrm{ch}}^{\mathrm{load}}<1$ fJ/bit for $P_{\mathrm{bias}}\sim100\ \mu$W and $V_{\mathrm{EOM}}^{\mathrm{load}}\sim0.5$ V. With $R_{\mathrm{load}}=500$ $\Omega$, the energy consumption is reduced owing to the high-speed operation. More generally, reducing $R_{\mathrm{load}}$ decreases the RC time constant and increases the bandwidth, whereas reducing $C_\mathrm{tot}$ simultaneously improves both speed and energy efficiency through Eq. \ref{load-resistor_charge}. Therefore, these results indicate that load-resistor-type OEO converters can achieve high-speed operation with low energy consumption by minimizing the device capacitance (e.g., by reducing pad and interconnect parasitics) and improving the EO efficiency. 

\paragraph{Current-injection type}
Here, the EOM is current-driven in carrier-injection mode, and the dynamic term is
\begin{equation}\label{current-injection_charge}
E_{\mathrm{ch}}^{\mathrm{inj}}=Q_{\mathrm{sw}}V_{\mathrm{EOM}}^{\mathrm{inj}}/4,
\end{equation}
where $Q_{\mathrm{sw}}\simeq\Delta I_{\mathrm{EOM}}\tau_{f}$ and $V_{\mathrm{EOM}}^{\mathrm{inj}}=\Delta I_{\mathrm{EOM}}R_{\mathrm{on}}$, with $\Delta I_{\mathrm{EOM}}$ the forward current swing, $\tau_{f}$ the effective carrier lifetime (fall time), and $R_{\mathrm{on}}$ the forward resistance. At $f_{\mathrm{mod}}=100$ Mb/s, we obtain $E_{\mathrm{opt\_PD}}=2.5$ pJ/bit and $E_{\mathrm{elect}}=4.5$ pJ/bit for $P_{\mathrm{PD}}=1.0$ mW, $\eta_{\mathrm{PD}}=0.9$ A/W, and $V_{\mathrm{PD}}=-2$ V. With $P_{\mathrm{bias}}\sim1$ mW (corresponding to RF OEO gain$\sim1$) and $f_{\mathrm{mod}}=100$ Mb/s, we obtain $E_{\mathrm{opt\_EOM}}=10$ pJ/bit. Taking $\Delta I_{\mathrm{EOM}}=1.0$ mA and $\tau_{f}=2.1$ ns gives $Q_{\mathrm{sw}}=2.1$ pC. Assuming $R_{\mathrm{on}}\approx100\ \Omega$ (so $V_{\mathrm{EOM}}^{\mathrm{inj}}\approx100$ mV), $E_{\mathrm{ch}}^{\mathrm{inj}}\approx50$ fJ/bit. With improved EO efficiency, the required $P_{\mathrm{bias}}$ can be reduced toward $E_{\mathrm{opt\_EOM}}\sim1$ pJ/bit. However, the modulation speed is fundamentally limited by $\tau_{f}$ (carrier recombination), which constrains the energy reduction at high symbol rates.

In both OEO schemes, $E_{\mathrm{OEO,bit}}$ is typically dominated by $E_{\mathrm{opt\_EOM}}$, whereas the attainable speed is RC-limited (load-resistor) or lifetime-limited (current-injection). Thus, reducing EOM bias power and device capacitance is key to lowering energy, while carrier-lifetime engineering is critical for accelerating current-injection operation. Heterogeneous co-integration of III-V (e.g., InP) or SiGe devices with silicon photonics can further reduce the operating energy of load-resistor-type OEO converters.

\section{Conclusion}
Monolithically integrated silicon-photonic OEO converters in load-resistor (high-speed variant) and current-injection (high-gain variant) configurations provide on-chip RF gain and reconfigurable nonlinear activation functions. The load-resistor device achieves a net RF gain at an operating data rate of 200 Mb/s with the $R_\mathrm{load}$ of 10 k$\Omega$. A higher-speed variant with $R_\mathrm{load}=500$ $\Omega$ shows clear eye openings up to 4 Gb/s, indicating the effective signal amplification and high-speed modulation capability. To the best of our knowledge, this is the first experimental demonstration of a monolithically integrated silicon-photonic OEO converter of the load-resistor type without a TIA showing reconfigurable nonlinear transfer and measurable on-chip RF OEO gain. In contrast, the current-injection configuration exhibits a stronger RF OEO gain and enhanced ER regeneration, attributed to its active carrier modulation. However, its RF bandwidth and operating speed are limited to about 100 Mb/s, primarily by the carrier recombination lifetime. Short-pulse measurements provide 3-dB bandwidths of 1.49 GHz, 160 MHz (load-resistor, $R_\mathrm{load}=500$ $\Omega$ and 10 k$\Omega$), and 76 MHz (current-injection), consistent with the RF data. These results indicate that the load-resistor-type OEO converters are suitable for high-speed and low-power-consumption ONNs that require broad bandwidth and optical gain, while the current-injection-type converters are well suited to optoelectronic computing and communication that process low-ER and low-power optical signals. OEO energy analysis indicates that reducing $C_{\mathrm{int}}$ and $P_{\mathrm{bias}}$ (e.g., via pad/parasitic minimization, higher EO efficiency, and heterogeneous co-integration) directly lowers the operating energies, charting a path to $<1$ pJ/bit for the load-resistor-type OEO converter. These foundry-compatible and CMOS-process-compatible OEO primitives thus offer a practical path to scalable and low-latency on-chip optoelectronic computing, RF photonic signal processing, and next-generation telecommunication systems, bridging the gap between photonics and electronics through monolithic silicon integration.

\section*{Methods}
The chips were monolithically fabricated at AMF on a silicon photonics platform. The spectra measurement of a stand-alone MRM and OEO converters (nonlinear activation functions) were performed by a wavelength-tunable CW laser and an optical power meter (Agilent 8164B). In RF OEO gain measurement, the RF input signals to the PD were prepared by a CW laser (Santec TSL-510), a signal generator (AWG Keyseight M9502A), and a Mach-Zehnder modulator (MZM, QAM(IQ) Transmitter BOX). The MZM is driven by programmed pseudo-random binary sequence (PRBS) patterns of $2^7-1$ bits at a data rate to yield an NRZ signal. The wavelength-tunable CW laser (Santec TSL-570) is incident on EOM and then the output signal is recorded by a sampling oscilloscope (Keysight 86100D Infiniium DCA-X). In the short-pulse response measurement, a pulse laser (Pritel, with a pulse width of 22 ps) is incident on the PD.

\section*{Acknowledgments}
This work was supported by JST-CREST under Grant JPMJCR21C3.

\section*{Author declarations}
\subsection*{Conflict of Interest}
The authors have no conflicts to disclose.

\section*{Data Availability}
The data that support the findings of this study are available
from the corresponding author upon reasonable request.

\section*{References}

\nocite{*}
\bibliography{paper}% Produces the bibliography via BibTeX.

\end{document}